\documentclass{optica-article}

\journal{opticajournal}

\articletype{Research Article}

\usepackage{siunitx}
\usepackage{physics}

\begin{document}

\title{Orchestrating time and color: a programmable source of high-dimensional entanglement}

\author{Laura Serino,\authormark{1,$\dag$,*} Werner Ridder,\authormark{1,$\dag$} Abhinandan Bhattacharjee,\authormark{1} Jano Gil-Lopez,\authormark{1} Benjamin Brecht,\authormark{1} and Christine Silberhorn\authormark{1}}

\address{\authormark{1}Paderborn University, Integrated Quantum Optics, Institute for Photonic Quantum Systems (PhoQS), Warburgerstr.\ 100, 33098 Paderborn, Germany\\
\authormark{$\dag$}The authors contributed equally to this work.}

\email{\authormark{*}laura.serino@upb.de}

\begin{abstract*}
High-dimensional encodings based on temporal modes (TMs) of photonic quantum states provide the foundations for a highly versatile and efficient quantum information science (QIS) framework. Here, we demonstrate a crucial building block for any QIS applications based on TMs: a programmable source of maximally entangled high-dimensional TM states. Our source is based on a parametric down-conversion process driven by a spectrally shaped pump pulse, which facilitates the generation of maximally entangled TM states with a well-defined dimensionality that can be chosen programmatically. We characterize the effective dimensionality of the generated states via measurements of second-order correlation functions and joint spectral intensities, demonstrating the generation of bi-photon TM states with a controlled dimensionality in up to 20 dimensions.
\end{abstract*}

\section{Introduction}
In recent years, the field of quantum communication has experienced remarkable progress, bringing us closer to the creation of a so-called quantum internet \cite{kimble08} which will exploit the fundamental properties of quantum particles to guarantee secure and efficient transmission of information. 
High-dimensional entangled states play a key role in these advancements \cite{yuan10}: their higher information capacity allows for significantly more efficient communication, and high-dimensional quantum cryptography protocols offer enhanced security \cite{sheridan10}. Photons emerge as a natural information carrier due to their inherent quantum nature and high-dimensional spatial and time-frequency degrees of freedom.

Of the high-dimensional photonic degrees of freedom, the spatial domain is arguably the most explored, due to the possibility to generate and manipulate states using only time-invariant operations \cite{berkhout10, huang18, mirhosseini15}. However, spatial encoding is incompatible with existing single-mode fiber networks and information are easily degraded by turbulence in free-space transmission \cite{cozzolino19}.

Encoding information in the time-frequency domain of photons overcomes these disadvantages by offering robust transmission both through optical fiber and free space. In this platform, information can be encoded in pulsed temporal modes (TMs) \cite{brecht15}, i.e., field-orthogonal wave-packet modes. A practical quantum communication framework based on TMs requires tools to generate, manipulate and detect high-dimensional TM states. The manipulation and detection of one or many single-photon TMs have been demonstrated through the so-called quantum pulse gate \cite{brecht14, ansari18a, serino23}, interferometric systems \cite{humphreys13} or a combination of phase modulators and pulse shapers \cite{lu18a, lu18b}. However, the realization of an optimal source of high-dimensional entangled TM states remains an open challenge.

An ideal source of time-frequency entanglement must generate bi-photon states with a well-defined dimensionality, i.e., states that live in a finite-dimensional Hilbert space \cite{brecht15}. A prime example of this are high-dimensional maximally entangled states, 
which facilitate entanglement-based quantum communication protocols such as high-dimensional device-independent quantum key distribution \cite{sheridan10}.
Moreover, the ability to programmatically reconfigure the dimensionality of the generated states can enhance the performance of a wide range of QIS applications, where different protocols and receivers may demand specific dimensionalities for efficient operation.
However, current implementations of high-dimensional TM sources typically meet only one of these essential requirements.
In dispersion-engineered parametric-down-conversion (PDC) sources \cite{graffitti20, morrison22, chiriano23}, high-dimensional entangled states are generated by precisely tailoring the phase-matching function of nonlinear crystals, preventing the dimensionality of the system from being modified after the fabrication process.
Gas-filled photonic crystal fibers are nonlinear materials whose dispersion properties can be altered even after fabrication, allowing for flexible dimensionality control of the generated entangled states \cite{finger17}; however, in these materials, the dispersion control is not accurate enough to guarantee a well-defined dimensionality.

Alternative reconfigurable schemes rely on spectral manipulation or filtering of broadband frequency-entangled photons applied through pulse shapers \cite{bernhard13, lingaraju19}, Hong-Ou-Mandel interference \cite{jin16}, or Fabry-P\'erot cavities after \cite{xie15, maltese20, chang21} or around \cite{ou99} the PDC source. 
These methods generate discrete frequency-bin-entangled states that can be scaled to higher dimensions by increasing the spectral bandwidth of the PDC source; however, these states are not maximally entangled due to the presence of residual frequency anti-correlations \cite{lu23} or uneven population of the frequency bins.

\begin{figure}
    \centering
    \includegraphics[]{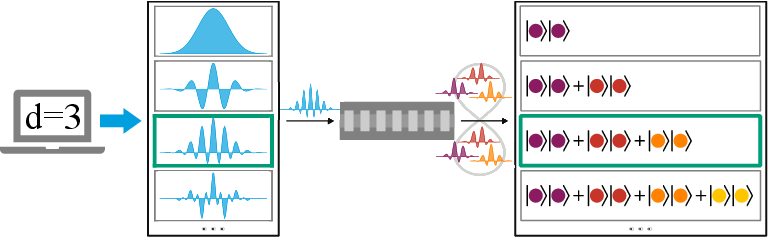}
    \caption{Working principle of the programmable source of high-dimensional entangled photons: the user-chosen dimensionality defines the pump spectrum for the PDC process, which determines the generated entangled state.}
    \label{fig:concept}
\end{figure}

Here, we demonstrate a programmable source of high-dimensional maximally entangled TM states (Fig.\ \ref{fig:concept}). Our source is based on a simple setup consisting of an initially spectrally decorrelated PDC process driven by a spectrally shaped pump pulse. Shaping the pump pulse into so-called cosine-kernel functions \cite{patera12} facilitates the generation of maximally entangled TM states with well-defined dimensionality within a fixed spectral bandwidth. Notably, pulse shaping does not require an overhaul of the experimental setup for accessing different dimensions. The effective dimensionality of the generated states is characterized via measurements of second-order correlation functions $g^{(2)}$, which link photon number statistics and modal properties \cite{christ11}. The modal structure is additionally analyzed via joint spectral intensities that directly probe the spectral distribution of the generated photon pairs, combined with the \textit{a priori} knowledge of the pump phase. We demonstrate the generation of bi-photon TM states with a controlled dimensionality starting from a fully decorrelated (unentangled) state and scaling up to a 20-dimensional entangled state.

\section{Theoretical background}
\begin{figure}
    \centering
    \includegraphics[]{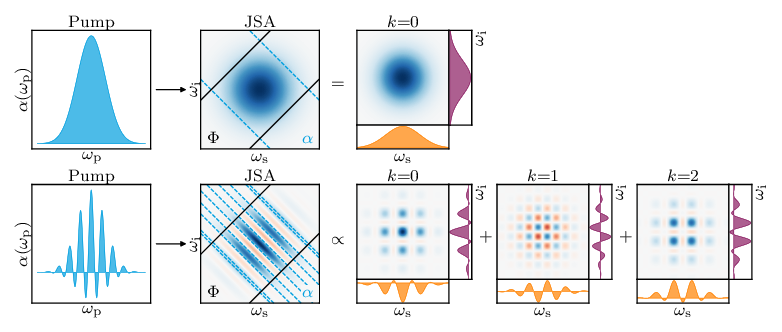}
    \caption{Visual representation of the Schmidt decomposition of the JSA of a PDC state generated by a pump shaped as a Gaussian (top) and second-order CK (bottom) function. From left to right: pump spectrum $\alpha(\omega_\mathrm{p})$, where $\omega_\mathrm{p}$ is the pump frequency; JSA, equal to the product of the pump function $\alpha(\omega_\mathrm{s} + \omega_\mathrm{i})$ (dashed contour) and phase-matching function $\Phi(\omega_\mathrm{s}, \omega_\mathrm{i})$ (solid contour); Schmidt modes and correspondent signal and idler functions. We note that the Schmidt modes in the bottom row have the same Schmidt coefficient $\sqrt\lambda=1/\sqrt{3}$.}
    \label{fig:schmidt}
\end{figure}

PDC is a process in which a \textit{pump} photon is converted into two lower-energy photons labeled \textit{signal} and \textit{idler}. This process must satisfy the requirements of energy conservation, described by the pump function $\alpha(\omega_\mathrm{s} + \omega_\mathrm{i})$ in terms of the signal and idler frequencies $\omega_\mathrm{s}$ and $\omega_\mathrm{i}$, respectively, and momentum conservation, given by the phase-matching function $\Phi(\omega_\mathrm{s}, \omega_\mathrm{i})$ which is fixed by material properties. The product of these two functions is the joint spectral amplitude (JSA)
\begin{equation}
\label{eq:jsa}
    f(\omega_\mathrm{s}, \omega_\mathrm{i}) = \alpha(\omega_\mathrm{s} + \omega_\mathrm{i}) \Phi(\omega_\mathrm{s}, \omega_\mathrm{i}) \,,
\end{equation}
which describes the correlations between the signal and idler photons as a function of their frequencies and, therefore, fully characterizes the PDC state \cite{law00} (Fig. \ref{fig:schmidt}).

The effective dimensionality of the Hilbert space in which the PDC state lives can be determined through a Schmidt decomposition of the corresponding JSA \cite{law00}. This process decomposes the JSA into sets of pairwise orthogonal modes
\begin{equation}
\label{eq:schmidt}
    f(\omega_\mathrm{s}, \omega_\mathrm{i}) = \sum_{k=0}^{r-1} \sqrt{\lambda_k} \psi^k(\omega_\mathrm{s}) \varphi^k(\omega_\mathrm{i}) \,,
\end{equation}
which are separable into signal ($\psi^k(\omega_\mathrm{s})$) and idler ($\varphi^k(\omega_\mathrm{i})$) complex spectral functions. The Schmidt coefficients $\sqrt{\lambda_k}$ indicate the relative amplitude of each mode pair and are normalized such that $\sum_{k} \lambda_k = 1$. The effective dimensionality of the PDC state is quantified by the Schmidt number $K = 1/(\sum_{k} \lambda_k^2)$, whereas the number of non-zero $\sqrt{\lambda_k}$ coefficients defines the Schmidt rank $r$. A finite value of $r$ implies that the generated signal and idler photons can be fully described by a finite-dimensional basis. In an experimental setting, this property allows one to fully characterize the PDC state through a finite number of measurements. If $K$=$r$, i.e., the Schmidt coefficients are uniformly distributed, then the photon pair is maximally entangled and can be described as $\ket{\Psi} = \sqrt{\lambda} \sum_{k=0}^{r-1} \ket{\psi^k} \ket{\varphi^k}$.

A Schmidt number of $K$=1 indicates a single-mode bi-photon state, i.e., a spectrally decorrelated JSA. This can be realized for instance through symmetric group-velocity matching in type-II PDC with a phase-matching function angled at \SI{45}{\degree} in the signal-idler frequency space \cite{harder13}. The positively-correlated phase-matching function will orthogonally intercept the pump function, which is always oriented at \SI{-45}{\degree} due to energy conservation. If the pump spectrum and phase-matching function are both Gaussian-shaped and have the same bandwidth, then the resulting JSA will be a fully separable circle in frequency space (Fig.\ \ref{eq:schmidt}a). 

In order to increase the dimensionality of the entangled state, one needs to modify the distribution of the Schmidt coefficients via appropriately tailoring the JSA. Eq.\ \ref{eq:jsa} entails that this can be achieved by manipulating either the phase-matching function or the pump spectrum. While both approaches are feasible and have been demonstrated, only the latter allows one to apply these changes ``on the fly'', i.e., programmatically and without any hardware modifications \cite{ansari18b}. For this reason, we focus on complex spectral shaping of the pump pulse \cite{weiner00} as the tool to tailor the PDC state.

We note that this technique fundamentally relies on achieving an initially decorrelated JSA which, in turn, requires an underlying positively-correlated phase-matching function. Experimentally, this can be realized in periodically poled potassium titanyl phosphate (ppKTP), which allows one to reach a phase-matching angle of approximately \SI{49}{\degree} in bulk and \SI{60}{\degree} in waveguides. The procedure for achieving decorrelation has been detailed in \cite{eckstein11, harder13, ansari18b} and, for this work, is described in the Supplementary Material.

One possibility to generate high-dimensional TM-entangled states is to shape the pump spectrum as Hermite-Gauss functions \cite{ansari18a}. However, while this method allows for a well-defined dimensionality of the PDC state, it does not produce maximally entangled states beyond a two-dimensional Bell state \cite{brecht15}. Moreover, the spectral bandwidth spanned by Hermite-Gauss functions grows wider for increasing orders, posing a practical challenge for realistic implementations with a limited available bandwidth.

A notable alternative involves shaping the pump spectrum into the so-called cosine-kernel (CK) modes \cite{patera12}, i.e., Gaussian spectra modulated by a cosine function, which are the frequency-space representation of a superposition of Gaussian time bins. A pump pulse shaped as the $n$-th order CK mode generates an ($n$+1)-dimensional maximally entangled state (Fig.\ \ref{fig:schmidt}). Unlike Hermite-Gauss modes, the spectral bandwidth spanned by CK modes is constant for all mode orders, offering a significant advantage for shaping methods similar to the one in this work, in which the pump spectrum is ``carved'' from an initial pulse with a fixed spectral bandwidth. CK modes present finer spectral features for increasing mode orders; consequently, the highest-order mode that can be realized experimentally, and thereby the highest achievable dimensionality, will then depend solely on the resolution of the shaping system. 

To verify the generation of high-dimensional maximally entangled states, one must characterize the modal structure of the PDC process \cite{law00}.
As mentioned above, this can be done by fully characterizing the JSA and directly performing a Schmidt decomposition to verify that the Schmidt coefficients are uniformly distributed. From the Schmidt coefficients, one can also calculate the Schmidt number $K$ to describe the effective dimensionality of the state. 
However, existing methods to fully characterize the JSA of PDC states (based on state tomography \cite{ansari18b} or spectral shearing interferometry \cite{davis20}) are extremely resource-expensive, as they require an additional nonlinear process or a complex interferometric setup.

In 2011, A. Christ \textit{et al.} \cite{christ11} demonstrated a resource-efficient method to probe $K$ using the normalized second-order correlation function $g^{(2)}$ of pulsed quantum light. This approach is suitable for a measurement system in which the detection window is longer than the pulse duration but shorter than the pulse period, which is the typical case in ultrafast quantum optics. In this regime, the value of $g^{(2)}$ indicates the integral of the second-order correlation function over one pulse. In the low-gain regime, the $g^{(2)}$ is connected to the Schmidt number $K$ by the simple relation \cite{christ11}
\begin{equation}
    g^{(2)} \approx 1 + \frac{1}{K} \,.
\end{equation}

Although $g^{(2)}$ measurements are an established tool to monitor the effective dimensionality of a PDC state, they are not sufficient to determine whether the state is maximally entangled, as a non-uniform distribution of Schmidt coefficients could also yield an integer value of $K$. Therefore, we look at $g^{(2)}$ measurements in combination with the joint spectral intensity (JSI) of the state. In contrast to the JSA, the JSI can be straightforwardly measured with a simple spectrometric setup requiring only a dispersive medium and time-resolved detection \cite{avenhaus09}. The JSI lacks the phase information necessary to correctly estimate the Schmidt coefficients; nevertheless, we can assume that the \emph{a priori} known spectral phase applied to the pump pulse is mapped to the JSA following the energy conservation condition, i.e., antidiagonally in the signal-idler frequency space. We can then apply this phase to the square root of the measured JSI to reconstruct the JSA, and confirm the maximally entangled condition of the PDC state via a Schmidt decomposition. This leads us to an estimate of the dimensionality of the system that we can compare to the effective dimensionality calculated from the $g^{(2)}$ to confirm the generation of a high-dimensional maximally entangled state.

\section{Experiment}

\begin{figure}
    \centering
    \includegraphics[]{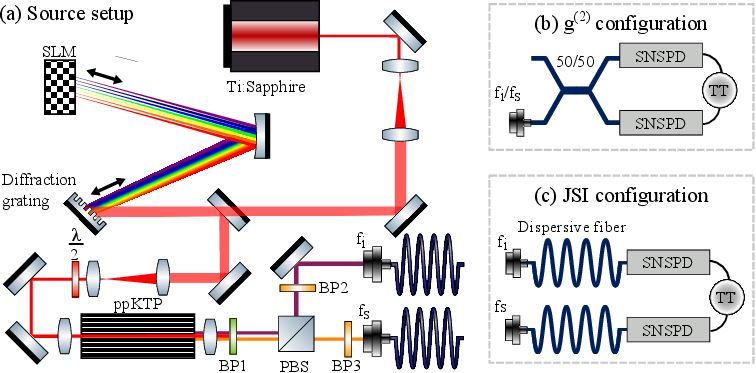}
    \caption{Schematic of the experimental setup. \textbf{a)} PDC source. The pump beam (red line) is generated by a Ti:Sapphire laser and is spectrally shaped by a 4-f waveshaper based on a spatial light modulator (SLM). The shaped beam is coupled into a periodically poled potassium titanyl phosphate (ppKTP) waveguide. The photon pair generated by the PDC process in the waveguide is isolated from the remaining pump light by a broad band pass filter (BP1). 
    The two photons are separated by a polarizing beam splitter (PBS), filtered again by narrowband filters BP2 and BP3 and then coupled into single-mode fibers. \textbf{b)} $g^{(2)}$-measurement configuration. One arm $f_{i/s}$ of the source setup is connected to a 50/50 fiber beam splitter. The photons in each output of the beam splitter are detected by a combination of a superconducting nanowire single-photon detectors (SNSPDs) and a time-tagging unit (TT). \textbf{c)} Measurement configuration for the time of flight (ToF) spectrograph. The frequencies of signal and idler photons are mapped to delays by dispersive fibers, and arrival-time correlations are detected by single-photon detectors and a time-tagging unit.}
    \label{fig:setup}
\end{figure} 

A schematic of the experiment is shown in Fig.\ \ref{fig:setup}a. Ultrashort laser pulses with a duration of \SI{150}{fs} are generated by a Ti:Sapphire laser with a repetition rate of 80 MHz and a spectrum centered at $\lambda_\mathrm{p} = \SI{758.7}{\nano\meter}$. The beam is directed to a folded-4-f waveshaper \cite{monmayrant10} with a resolution of approximately \SI{10}{\giga\hertz} to prepare the pump modes via amplitude and phase shaping. 
The pump beam is then coupled into an 8-mm-long periodically poled potassium titanyl phosphate (ppKTP) waveguide by AdvR Inc. with a poled length of \SI{1.5}{\mm} and a poling period of \SI{117}{\micro\meter}. 
The type-II PDC process inside the poled waveguide generates pairs of orthogonally polarized photons with central wavelengths $\lambda_\mathrm{0,s} = \SI{1511}{\nm}$ and $\lambda_\mathrm{0,i} = \SI{1524}{\nm}$ for signal and idler, respectively. 

A band-pass filter centered at \SI{1538}{\nm} with an acceptance bandwidth of \SI{82}{\nm} (Semrock FF01-1538/82) filters out the remaining pump light and most of the fluorescence light coming from the waveguide. 
The photon pair is then separated by a polarizing beam splitter (PBS) into the vertically polarized signal and the horizontally polarized idler.
To further filter out fluorescence from the waveguide, both arms contain a narrow band-pass filter with a full-width-half-maximum of \SI{7}{\nm} angle-tuned to match the centers of the signal and idler spectra. The filter bandwidth is very close to the spectral bandwidth of the photons (approximately \SI{7}{\nm} and \SI{5}{\nm} for signal and idler, respectively) as it allows for significantly suppressing fluorescence noise while transmitting most of the PDC photons. 
After filtering, the photons are coupled into single-mode fibers and detected by two superconducting nanowire single-photon detectors (SNSPDs) with an efficiency of 80\%.

We estimate the optical losses of the setup through the Klyshko efficiency \cite{klyshko80}, defined as $\eta_{\mathrm{s (i)}}=p_\mathrm{coinc}/p_\mathrm{i (s)}$ for the signal (idler) arm, where $p_\mathrm{coinc}$ is the probability to detect a coincidence between both arms and $p_\mathrm{i (s)}$ is the probability to have a click in the idler (signal) arm. Effectively, the Klyshko efficiency measures the probability of detecting one photon from a generated photon pair, given that the other photon has been detected. We measure $\eta_{\mathrm{s}}\approx25\%$ and $\eta_{\mathrm{i}}\approx25\%$, where the optical losses are mostly due to the fiber couplings. 

For the $g^{(2)}$ measurements, one of the fiber out-couplings is connected to a fiber beam splitter, each output of which is detected by an SNSPD (Fig.\ \ref{fig:setup}b). Photon arrival times are recorded by a Swabian Instruments time-tagging unit, and time filtering is performed to suppress the fluorescence counts. Using the $g^{(2)}$ value as reference, we optimize the bandwidth and chirp (quadratic phase coefficient) parameters via the waveshaper to obtain the initial decorrelated PDC state from a Gaussian-shaped pump (see Supplementary Material for details).

We measure the JSI using a time-of-flight (ToF) spectrograph \cite{avenhaus09} (Fig.\ \ref{fig:setup}c). In this configuration, a dispersive fiber with a chromatic dispersion coefficient of \SI{-418}{\ps/\nm} is added to each of the PDC output arms to map frequencies to delays. The arrival times of the photons are then measured by two SNSPDs connected to a time-tagging unit. The pulse duration is short enough not to affect this measurement. The correlation between the arrival times of paired signal and idler is used to reconstruct the JSI.

\section{Results and discussion}
\begin{figure}
    \centering
    \includegraphics[]{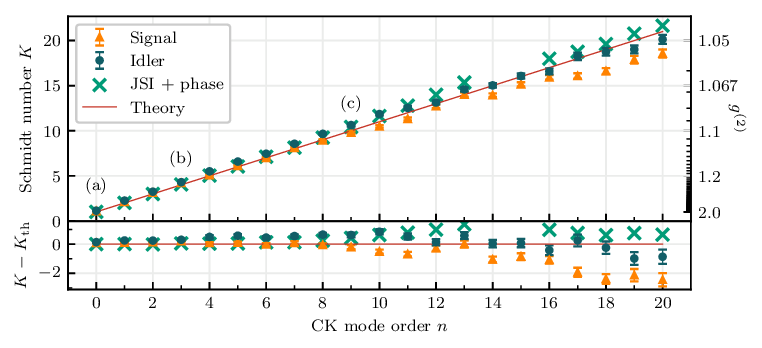}
    \caption{Experimental Schmidt number $K$ for each CK mode $n$ obtained from $g^{(2)}$ measurements in the signal and idler arms (orange and blue points) and values reconstructed from the deconvolved measured JSI combined with the known pump phase (green cross), compared to the theoretical values $K_\mathrm{th}$ (red line). The marked points correspond to the CK modes analyzed in Fig.\ \ref{fig:results_jsa}.}
    \label{fig:results_g2}
\end{figure}

\begin{figure}
    \centering
    \includegraphics[]{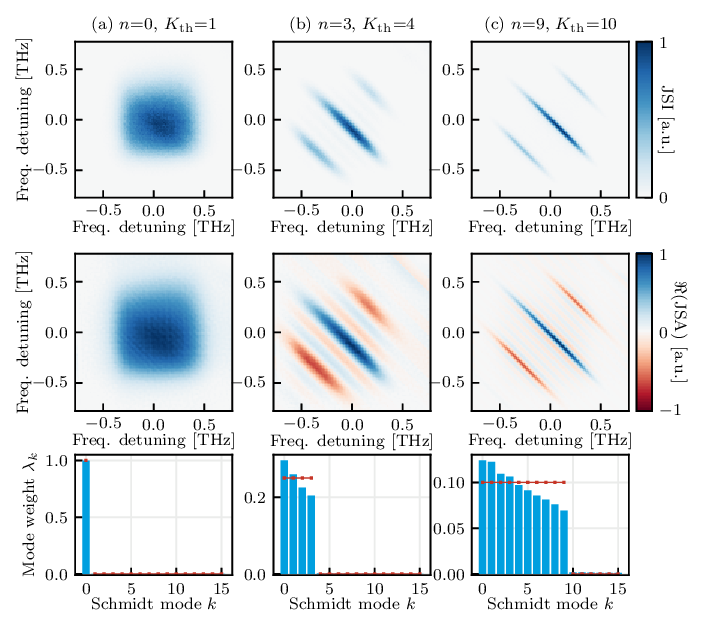}
    \caption{For three different CK mode orders $n$: measured JSI (first row), JSA reconstructed from the deconvolved measured JSI combined with the known pump phase (second row), and resulting distribution of the Schmidt weights $\lambda_k$ (blue bars) for each Schmidt mode $k$, compared to the expected weights for a maximally entangled state (red line).}
    \label{fig:results_jsa}
\end{figure}

Fig.\ \ref{fig:results_g2} shows the measured effective dimensionality $K$ of the PDC states as a function of the programmed CK order $n$ compared to the theoretical values $K_\mathrm{th}$. The experimental dimensionality, characterized via $g^{(2)}$ measurements of the signal and idler arms separately, grows linearly with $n$, confirming that the proposed system can successfully allow the user to programmatically select the effective dimensionality of the PDC state in up to 20 dimensions.

The initial values for the signal and idler arms, $g^{(2)}_{0,\mathrm{s}} = 1.93 \pm 0.01$ and $g^{(2)}_{0,\mathrm{i}} = 1.88 \pm 0.01$, correspond to a Schmidt number $K_{0,\mathrm{s}} = 1.08 \pm 0.01$ and $K_{0,\mathrm{i}} = 1.14 \pm 0.01$, respectively. These values are slightly larger than the expected fully decorrelated state value of $K_{0,\mathrm{th}} = 1$, primarily due to non-ideal phase-matching conditions: the phase-matching function of the ppKTP waveguide, shaped as a sinc at an angle of \SI{31}{\degree} with respect to the signal axis, deviates from the ideal condition of a perfectly diagonal Gaussian phase-matching.
This non-ideal configuration leads to unwanted correlations between signal and idler frequencies, introducing a small multi-mode component to the PDC state.
While these imperfections could be mitigated with the use of an apodized waveguide with a Gaussian phase-matching spectrum \cite{dosseva16, graffitti17} engineered to have a perfect positive frequency correlation between signal and idler, their impact in this source is limited and is further reduced by spectral filters, which cut out the side lobes of the sinc-shaped phase-matching function. 

Above $K=18$, one can notice a higher discrepancy between the theoretical and experimental dimensionality. This is a consequence of the limited resolution of the waveshaper, which fails to perfectly shape the narrow features of high-order CK functions. As a matter of fact, the CK function of order $n=17$ has a main peak with a spectral width of approximately \SI{30}{\GHz}, only three times larger than the resolution of the waveshaper. This issue can be easily addressed by using an improved waveshaper: increasing the resolution by only a factor 2 would allow one to generate CK functions up to order $n=32$ before reaching the shaping limits.

Fig.\ \ref{fig:results_jsa} (a), (b) and (c) show, for three different CK orders, the measured JSI (first row), the JSA reconstructed by applying the known pump phase to the measured JSI (second row) and the weights $\lambda_k$ obtained from the Schmidt decomposition of the JSA (last row). We note that the limited resolution of the ToF spectrograph introduced an artificial broadening of the fine spectral features of high-order CK spectra; therefore, to eliminate these artifacts, we performed a deconvolution of the direct JSI measurements using the measured point-spread function of the ToF setup (see Supplementary Material for a detailed explanation).

The resulting Schmidt weights distribution (in blue) presents a clear edge at $k=K_\mathrm{th}$, as expected from a maximally entangled state (in red). 
The zero value of the Schmidt weights beyond $K_\mathrm{th}$ indicates that the presented source generates states with a finite dimensionality, i.e., containing a finite number of Schmidt modes. 
The main discrepancy with the theory lies in the uneven value of the coefficients, which we attribute to the spectral distortions induced by the narrowband filters in each arm.
As one can observe in the square edges in Fig.\ \ref{fig:results_jsa}(a), the filters have a slightly narrower bandwidth than the generated photons and, as a consequence, introduce spectral distortions in the PDC state, which result in uneven additional optical losses for different Schmidt modes. 
Despite this drawback, narrow filtering is beneficial for $g^{(2)}$ measurements as it limits fluorescence noise, which would otherwise artificially increase the detected single counts and lead to a less accurate assessment of the $g^{(2)}$ value of the source. Contrarily, since JSI measurements are not as significantly affected by fluorescence, the effect of the induced distortions becomes noticeable. This is especially evident in the signal arm, which has a wider spectrum than the idler due to the sub-optimal phase-matching angle.
Nevertheless, the unevenness of the Schmidt weights is small enough not to significantly affect $K$, as indicated by the Schmidt number $K$ calculated from this distribution (the green cross in Fig.\ \ref{fig:results_g2}) being very close to the theoretical value $K_\mathrm{th}$. Furthermore, the Schmidt weights can be directly addressed by tailoring the pump spectrum, allowing one to make their distribution more even if necessary.

We refer the reader to the Supplementary Material for a more detailed analysis of the causes of experimental imperfections and possible solutions.
We note that the highlighted challenges are technical, rather than fundamental limitations, and can be readily addressed, e.g., by employing a dispersion-engineered waveguide, a higher-resolution waveshaper, and spectral filters with an optimally matched bandwidth.

\section{Conclusion}
We demonstrated the generation of time-frequency-entangled photons with a programmable dimensionality ranging from 1 (decorrelated state) up to 20 dimensions, selected via straightforward spectral shaping of the pump pulse.
By analyzing the measured joint spectral intensity alongside the \textit{a priori} knowledge of the pump phase, we inferred a Schmidt coefficients distribution compatible with high-dimensional maximally entangled states. 
While our results and the theoretical predictions strongly suggest maximal time-frequency entanglement, a rigorous proof will necessitate a careful phase-sensitive characterization of the generated PDC states. 
This can be accomplished through high-dimensional state tomography, facilitated, e.g., by a so-called multi-output quantum pulse gate \cite{serino23}.

The performance of the current system is limited solely by technical constraints, which can be substantially enhanced by incorporating state-of-the-art experimental components. These improvements will facilitate the generation of maximally entangled states in more than 40 dimensions. Beyond dimensional control, manipulating the pump spectrum offers the potential not only to program the state dimensionality, but also to finely adjust the Schmidt weights themselves, thereby achieving an unprecedented degree of entanglement control.

The ongoing development of our source and the integration of cutting-edge components will pave the way for groundbreaking applications in quantum information processing, quantum communication, and quantum metrology, establishing our platform as a promising resource for advancing the capabilities of quantum technologies.

\begin{backmatter}
\bmsection{Funding}
This research was supported by the EU H2020 QuantERA ERA-NET Cofund in Quantum Technologies project QuICHE, and by the Deutsche Forschungsgemeinschaft (DFG, German Research Foundation) – SFB-Geschäftszeichen TRR142/3-2022 – Projektnummer 231447078; Project C07.

\bmsection{Acknowledgments}
The authors thank V. Ansari for helpful discussions.

\bmsection{Disclosures}
The authors declare no conflicts of interest.

\bmsection{Data availability}
Data underlying the results presented in this paper are not publicly available at this time but may be obtained from the authors upon reasonable request.

\bmsection{Supplemental document}
See Supplement 1 for supporting content. 

\end{backmatter}

\bibliography{bibliography}

\begin{thebibliography}{10}
\newcommand{\enquote}[1]{``#1''}

\bibitem{kimble08}
H.~J. Kimble, \enquote{The quantum internet,} {\protect\JournalTitle{Nature}} \textbf{453}, 1023--1030 (2008).

\bibitem{yuan10}
Z.-S. Yuan, X.-H. Bao, C.-Y. Lu, \emph{et~al.}, \enquote{Entangled photons and quantum communication,} {\protect\JournalTitle{Physics Reports}} \textbf{497}, 1--40 (2010).

\bibitem{sheridan10}
L.~Sheridan and V.~Scarani, \enquote{Security proof for quantum key distribution using qudit systems,} {\protect\JournalTitle{Physical Review A}} \textbf{82} (2010).

\bibitem{berkhout10}
G.~C.~G. Berkhout, M.~P.~J. Lavery, J.~Courtial, \emph{et~al.}, \enquote{Efficient sorting of orbital angular momentum states of light,} {\protect\JournalTitle{Phys. Rev. Lett.}} \textbf{105}, 153601 (2010).

\bibitem{huang18}
K.~Huang, H.~Liu, S.~Restuccia, \emph{et~al.}, \enquote{Spiniform phase-encoded metagratings entangling arbitrary rational-order orbital angular momentum,} {\protect\JournalTitle{Light: Science \& Applications}} \textbf{7}, 17156--17156 (2018).

\bibitem{mirhosseini15}
M.~Mirhosseini, O.~S. Maga{\~{n}}a-Loaiza, M.~N. O'Sullivan, \emph{et~al.}, \enquote{High-dimensional quantum cryptography with twisted light,} {\protect\JournalTitle{New Journal of Physics}} \textbf{17}, 033033 (2015).

\bibitem{cozzolino19}
D.~Cozzolino, B.~Da~Lio, D.~Bacco, and L.~K. Oxenløwe, \enquote{High-dimensional quantum communication: Benefits, progress, and future challenges,} {\protect\JournalTitle{Advanced Quantum Technologies}} \textbf{2}, 1900038 (2019).

\bibitem{brecht15}
B.~Brecht, D.~V. Reddy, C.~Silberhorn, and M.~G. Raymer, \enquote{Photon temporal modes: A complete framework for quantum information science,} {\protect\JournalTitle{Phys. Rev. X}} \textbf{5}, 041017 (2015).

\bibitem{brecht14}
B.~Brecht, A.~Eckstein, R.~Ricken, \emph{et~al.}, \enquote{Demonstration of coherent time-frequency schmidt mode selection using dispersion-engineered frequency conversion,} {\protect\JournalTitle{Phys. Rev. A}} \textbf{90}, 030302(R) (2014).

\bibitem{ansari18a}
V.~Ansari, E.~Roccia, M.~Santandrea, \emph{et~al.}, \enquote{Heralded generation of high-purity ultrashort single photons in programmable temporal shapes,} {\protect\JournalTitle{Opt. Express}} \textbf{26}, 2764--2774 (2018).

\bibitem{serino23}
L.~Serino, J.~Gil-Lopez, M.~Stefszky, \emph{et~al.}, \enquote{Realization of a multi-output quantum pulse gate for decoding high-dimensional temporal modes of single-photon states,} {\protect\JournalTitle{PRX Quantum}} \textbf{4}, 020306 (2023).

\bibitem{humphreys13}
P.~C. Humphreys, B.~J. Metcalf, J.~B. Spring, \emph{et~al.}, \enquote{Linear optical quantum computing in a single spatial mode,} {\protect\JournalTitle{Phys. Rev. Lett.}} \textbf{111}, 150501 (2013).

\bibitem{lu18a}
H.-H. Lu, J.~M. Lukens, N.~A. Peters, \emph{et~al.}, \enquote{Electro-optic frequency beam splitters and tritters for high-fidelity photonic quantum information processing,} {\protect\JournalTitle{Phys. Rev. Lett.}} \textbf{120}, 030502 (2018).

\bibitem{lu18b}
H.-H. Lu, J.~M. Lukens, N.~A. Peters, \emph{et~al.}, \enquote{Quantum interference and correlation control of frequency-bin qubits,} {\protect\JournalTitle{Optica}} \textbf{5}, 1455--1460 (2018).

\bibitem{graffitti20}
F.~Graffitti, P.~Barrow, A.~Pickston, \emph{et~al.}, \enquote{Direct generation of tailored pulse-mode entanglement,} {\protect\JournalTitle{Physical Review Letters}} \textbf{124}, 1--6 (2020).

\bibitem{morrison22}
C.~L. Morrison, F.~Graffitti, P.~Barrow, \emph{et~al.}, \enquote{{Frequency-bin entanglement from domain-engineered down-conversion},} {\protect\JournalTitle{APL Photonics}} \textbf{7}, 066102 (2022).

\bibitem{chiriano23}
F.~Chiriano, J.~Ho, C.~L. Morrison, \emph{et~al.}, \enquote{Hyper-entanglement between pulse modes and frequency bins,} {\protect\JournalTitle{Opt. Express}} \textbf{31}, 35131--35142 (2023).

\bibitem{finger17}
M.~A. Finger, N.~Y. Joly, P.~S.~J. Russell, and M.~V. Chekhova, \enquote{Characterization and shaping of the time-frequency schmidt mode spectrum of bright twin beams generated in gas-filled hollow-core photonic crystal fibers,} {\protect\JournalTitle{Physical Review A}} \textbf{95}, 1--10 (2017).

\bibitem{bernhard13}
C.~Bernhard, B.~Bessire, T.~Feurer, and A.~Stefanov, \enquote{{Shaping frequency-entangled qudits},} {\protect\JournalTitle{Physical Review A - Atomic, Molecular, and Optical Physics}} \textbf{88}, 1--4 (2013).

\bibitem{lingaraju19}
N.~B. Lingaraju, H.-H. Lu, S.~Seshadri, \emph{et~al.}, \enquote{{Quantum frequency combs and Hong–Ou–Mandel interferometry: the role of spectral phase coherence},} {\protect\JournalTitle{Optics Express}} \textbf{27}, 38683 (2019).

\bibitem{jin16}
R.~B. Jin, R.~Shimizu, M.~Fujiwara, \emph{et~al.}, \enquote{Simple method of generating and distributing frequency-entangled qudits,} {\protect\JournalTitle{Quantum Science and Technology}} \textbf{1} (2016).

\bibitem{xie15}
Z.~Xie, T.~Zhong, S.~Shrestha, \emph{et~al.}, \enquote{{Harnessing high-dimensional hyperentanglement through a biphoton frequency comb},} {\protect\JournalTitle{Nature Photonics}} \textbf{9}, 536--542 (2015).

\bibitem{maltese20}
G.~Maltese, M.~I. Amanti, F.~Appas, \emph{et~al.}, \enquote{{Generation and symmetry control of quantum frequency combs},} {\protect\JournalTitle{npj Quantum Information}} \textbf{6} (2020).

\bibitem{chang21}
K.-C. Chang, X.~Cheng, M.~C. Sarihan, \emph{et~al.}, \enquote{{648 Hilbert-space dimensionality in a biphoton frequency comb: entanglement of formation and Schmidt mode decomposition},} {\protect\JournalTitle{npj Quantum Information}} \textbf{7}, 48 (2021).

\bibitem{ou99}
Z.~Y. Ou and Y.~J. Lu, \enquote{{Cavity enhanced spontaneous parametric down-conversion for the prolongation of correlation time between conjugate photons},} {\protect\JournalTitle{Physical Review Letters}} \textbf{83}, 2556--2559 (1999).

\bibitem{lu23}
H.-H. Lu, M.~Liscidini, A.~L. Gaeta, \emph{et~al.}, \enquote{{Frequency-bin photonic quantum information},} {\protect\JournalTitle{Optica}} \textbf{10}, 1655 (2023).

\bibitem{patera12}
G.~Patera, C.~Navarrete-Benlloch, G.~J. de~Valc{\'a}rcel, and C.~Fabre, \enquote{Quantum coherent control of highly multipartite continuous-variable entangled states by tailoring parametric interactions,} {\protect\JournalTitle{The European Physical Journal D}} \textbf{66}, 241 (2012).

\bibitem{christ11}
A.~Christ, K.~Laiho, A.~Eckstein, \emph{et~al.}, \enquote{Probing multimode squeezing with correlation functions,} {\protect\JournalTitle{New Journal of Physics}} \textbf{13}, 033027 (2011).

\bibitem{law00}
C.~K. Law, I.~A. Walmsley, and J.~H. Eberly, \enquote{Continuous frequency entanglement: Effective finite hilbert space and entropy control,} {\protect\JournalTitle{Phys. Rev. Lett.}} \textbf{84}, 5304--5307 (2000).

\bibitem{harder13}
G.~Harder, V.~Ansari, B.~Brecht, \emph{et~al.}, \enquote{An optimized photon pair source for quantum circuits,} {\protect\JournalTitle{Opt. Express}} \textbf{21}, 13975--13985 (2013).

\bibitem{ansari18b}
V.~Ansari, J.~M. Donohue, M.~Allgaier, \emph{et~al.}, \enquote{Tomography and purification of the temporal-mode structure of quantum light,} {\protect\JournalTitle{Phys. Rev. Lett.}} \textbf{120}, 213601 (2018).

\bibitem{weiner00}
A.~M. Weiner, \enquote{{Femtosecond pulse shaping using spatial light modulators},} {\protect\JournalTitle{Review of Scientific Instruments}} \textbf{71}, 1929--1960 (2000).

\bibitem{eckstein11}
A.~Eckstein, A.~Christ, P.~J. Mosley, and C.~Silberhorn, \enquote{{Highly efficient single-pass source of pulsed single-mode twin beams of light},} {\protect\JournalTitle{Physical Review Letters}} \textbf{106}, 1--4 (2011).

\bibitem{davis20}
A.~O.~C. Davis, V.~Thiel, and B.~J. Smith, \enquote{{Measuring the quantum state of a photon pair entangled in frequency and time},} {\protect\JournalTitle{Optica}} \textbf{7}, 1317 (2020).

\bibitem{avenhaus09}
M.~Avenhaus, A.~Eckstein, P.~J. Mosley, and C.~Silberhorn, \enquote{Fiber-assisted single-photon spectrograph,} {\protect\JournalTitle{Opt. Lett.}} \textbf{34}, 2873--2875 (2009).

\bibitem{monmayrant10}
A.~Monmayrant, S.~Weber, and B.~Chatel, \enquote{Phd tutorial: A newcomer's guide to ultrashort pulse shaping and characterization,} {\protect\JournalTitle{Journal of Physics B-atomic Molecular and Optical Physics - J PHYS-B-AT MOL OPT PHYS}} \textbf{43} (2010).

\bibitem{klyshko80}
D.~N. Klyshko, \enquote{{Use of Two-Photon Light for Absolute Calibration of Photoelectric Detectors.}} {\protect\JournalTitle{Soviet journal of quantum electronics}} \textbf{7}, 1112--1116 (1980).

\bibitem{dosseva16}
A.~Dosseva, L.~Cincio, and A.~M. Bra\ifmmode~\acute{n}\else \'{n}\fi{}czyk, \enquote{Shaping the joint spectrum of down-converted photons through optimized custom poling,} {\protect\JournalTitle{Phys. Rev. A}} \textbf{93}, 013801 (2016).

\bibitem{graffitti17}
F.~Graffitti, D.~Kundys, D.~T. Reid, \emph{et~al.}, \enquote{{Pure down-conversion photons through sub-coherence-length domain engineering},} {\protect\JournalTitle{Quantum Science and Technology}} \textbf{2} (2017).

\end{thebibliography}


\begin{thebibliography}{1}
\newcommand{\enquote}[1]{``#1''}

\bibitem{richardson72}
W.~H. Richardson, \enquote{Bayesian-based iterative method of image restoration$\ast$,} {\protect\JournalTitle{J. Opt. Soc. Am.}} \textbf{62}, 55--59 (1972).

\bibitem{brecht14}
B.~Brecht, A.~Eckstein, R.~Ricken, \emph{et~al.}, \enquote{Demonstration of coherent time-frequency schmidt mode selection using dispersion-engineered frequency conversion,} {\protect\JournalTitle{Phys. Rev. A}} \textbf{90}, 030302(R) (2014).

\bibitem{ansari18a}
V.~Ansari, E.~Roccia, M.~Santandrea, \emph{et~al.}, \enquote{Heralded generation of high-purity ultrashort single photons in programmable temporal shapes,} {\protect\JournalTitle{Opt. Express}} \textbf{26}, 2764--2774 (2018).

\bibitem{serino23}
L.~Serino, J.~Gil-Lopez, M.~Stefszky, \emph{et~al.}, \enquote{Realization of a multi-output quantum pulse gate for decoding high-dimensional temporal modes of single-photon states,} {\protect\JournalTitle{PRX Quantum}} \textbf{4}, 020306 (2023).

\end{thebibliography}

\end{document}


\maketitle

\section{Cosine-kernel functions in the frequency domain}
\begin{figure}
    \centering
    \includegraphics{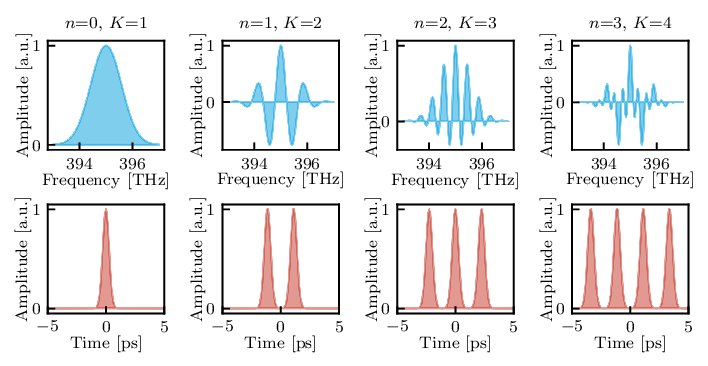}
    \caption{First four cosine-kernel (CK) functions expressed as a function of frequency (top) and time (bottom).}
    \label{fig:ck-fft}
\end{figure}

The cosine-kernel function of order $n$ in the frequency domain (Fig. \ref{fig:ck-fft}) represents a sequence of $K = n+1$ Gaussian-shaped time bins centered at time $t=0$ and separated by $\Delta t$, described in terms of frequency $\nu$:
\begin{equation}
    \mathrm{CK}_n(\nu) = \sum_{k=0}^{K-1} {\frac{1}{\sqrt{K}} \frac{1}{\sqrt{2 \pi \sigma^2}} e^{-\frac{(\nu - \nu_0)^2}{2 \sigma^2}} e^{i \, 2 \pi (\nu-\nu_0) \Delta t \, (k-\frac{K-1}{2})} }\,,
\end{equation}
where $\nu_0$ is the chosen central frequency, and $\sigma = \Delta \nu \,/ 2\sqrt2\ln2$, with $\Delta \nu$ being the chosen spectral full-width-half-maximum. After some simple calculations, this expression simplifies to
\begin{equation}
    \mathrm{CK}_n(\nu) = \frac{1}{\sqrt{K}} \frac{1}{\sqrt{2 \pi \sigma^2}} e^{-\frac{(\nu - \nu_0)^2}{2 \sigma^2}} \frac{\sin(\pi (\nu-\nu_0) \Delta t\, K)}{\sin(\pi (\nu-\nu_0) \Delta t)}\,,
\end{equation}
which describes a Gaussian envelope modulated by an oscillating function. These oscillations result in sharp peaks at frequencies $\nu_m = \nu_0 + m/\Delta t$ with null-to-null bandwidth $\delta\nu = 2/K \Delta t$.

Spectral shaping using CK functions is convenient for the shaping system employed in the current experimental setup: the Gaussian envelope limits the spectral bandwidth equally for all CK orders, and the spectral phase oscillates only between the two constant values 0 and $\pi$, which can be shaped more precisely than rapidly varying phases.

Describing CK modes as a sequence of $K$ time bins intuitively explains why a CK-shaped pump generates $K$-dimensional entangled PDC states. 
Moreover, one could take this technique one step further and assign different weights $\lambda_k$ to time bins, effectively tailoring the Schmidt coefficients and gaining full control over the modal structure of the entangled state.
Although this approach results in a complex-valued pump spectrum in which the spectral phase can assume any value between 0 and $2\pi$, this can still be generated by the current spectral shaping setup as long as the phase modulations are not excessively steep. This spectrum remains within the same Gaussian envelope, corresponding to the Fourier transform of a single time bin.

\section{Achieving a decorrelated PDC state}
The first step in the optimization of our source consists in finding the pump spectrum and spectral phase that facilitate the generation of a decorrelated PDC state (i.e., Schmidt number $K$=1). We expect this optimal pump shape to be a Gaussian spectrum with a bandwidth matching that of the phase-matching function. Additionally, the optical components in the setup will mostly introduce group-velocity dispersion in the pump, corresponding to a second-order phase coefficient (``chirp'') that must be compensated for by the waveshaper to achieve a fully decorrelated state. 
For this reason, we measure $g^{(2)}$ as a function of different values of the full-width-at-half-maximum (FWHM) and chirp correction, and select the ones that minimize $K$ (i.e., maximize the $g^{(2)}$).

\begin{figure}[h]
    \centering
    \includegraphics{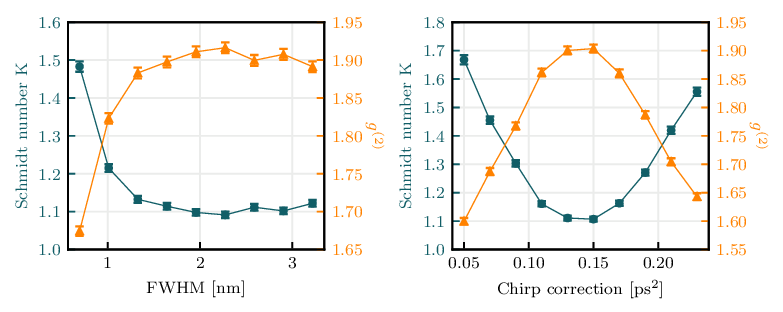}
    \caption{Schmidt number $K$ (blue circles) and $g^{(2)}$ (orange triangles) measured in the signal arm as a function of the FWHM of the pump spectrum (left) and of the chirp correction (right).}
    \label{fig:param-scan}
\end{figure}

The results of this characterization are shown in Fig. \ref{fig:param-scan}. One can notice that, above a FWHM of \SI{2}{\nm}, $K$ saturates because we run into the edges of the narrowband filters in the PDC arms (see Fig. 3 in the main text). 
We choose a bandwidth of \SI{2.6}{\nm} (\SI{1.3}{\tera\hertz}) which, after minor optimizations, yields $K = 1.08 \pm 0.01$ ($g^{(2)} = 1.93 \pm 0.01$) in the signal arm. 
This bandwidth is close to the phase-matching bandwidth of approximately \SI{1}{\tera\hertz}, and the discrepancy is to be attributed to the sub-optimal phase-matching angle, which is closer to \SI{29}{\degree} than to the ideal value of \SI{45}{\degree}. 
The optimal chirp correction, of approximately \SI{0.15}{\ps^2}, is a reasonable value to compensate for the chirp introduced by the setup.

\section{JSA reconstruction}
\begin{figure}
    \centering
    \includegraphics{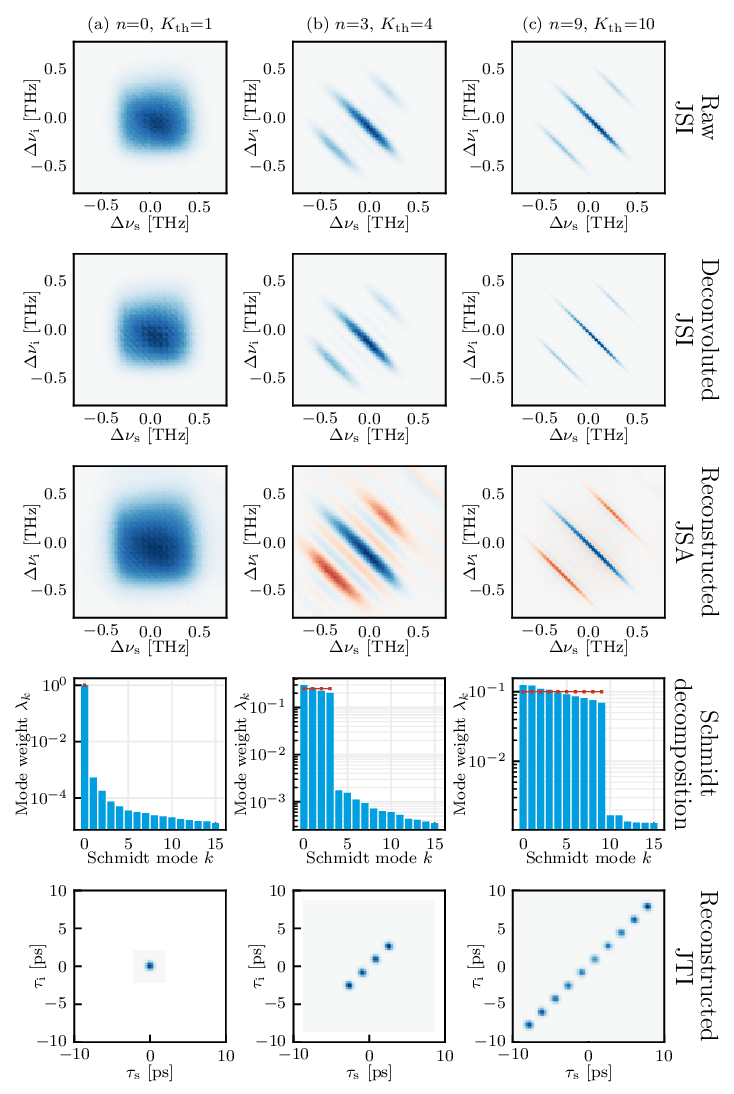}
    \caption{Reconstruction process of the JSA for three different mode orders. From top to bottom: ``raw'' JSI measurement, deconvoluted JSI, reconstructed JSA (taking the square root of the JSI and applying the known pump phase), mode weights $\lambda_k$ resulting from the Schmidt decomposition, reconstructed JTI (as squared modulus of the Fourier transform of the JSA).}
    \label{fig:jsa_rec}
\end{figure}

Fig. \ref{fig:jsa_rec} illustrates the JSA reconstruction process for three different mode orders. The first row shows the raw JSI measurement which, for high orders, is visibly blurred by the low effective resolution of the time-of-flight (ToF) spectrograph and, therefore, not representative of the real PDC state. To limit this effect, we measure the point-spread function (PSF) of the ToF setup and perform a Lucy-Richardson deconvolution \cite{richardson72} to retrieve the higher-resolution JSI shown in the second row. We note that this deconvolution algorithm does not require any information other than the blurred image and the measured PSF. At low CK orders, the deconvoluted JSI is essentially identical to the raw measurement; however, at high orders, the deconvolution brings back the distinctive fine features of the CK modes.   

To reconstruct the JSA, we take the square root of the deconvoluted JSI and we apply the known pump phase diagonally (centered on the maximum of the JSI), obtaining the image in the third row. The pump phase consists of a sequence of phase jumps between 0 (blue) and $\pi$ (red). We note that the phase jumps perfectly match the zero-intensity points in the JSI, confirming that the deconvoluted JSI is close to the real one. Then, we perform a Schmidt decomposition of the reconstructed JSA to retrieve the Schmidt coefficients $\sqrt{\lambda_k}$ which describe the modal structure of the generated PDC state. The distribution of the corresponding mode weights $\lambda_k$ for the analysed cases is shown in the fourth row. Since the Schmidt number derived this way matches the one calculated from the $g^{(2)}$ measurements, we can assume that the reconstructed JSA is representative of the real one and, therefore, that we can examine the modal structure of the generated PDC state from the calculated mode weights. 

\begin{figure}
    \centering
    \includegraphics{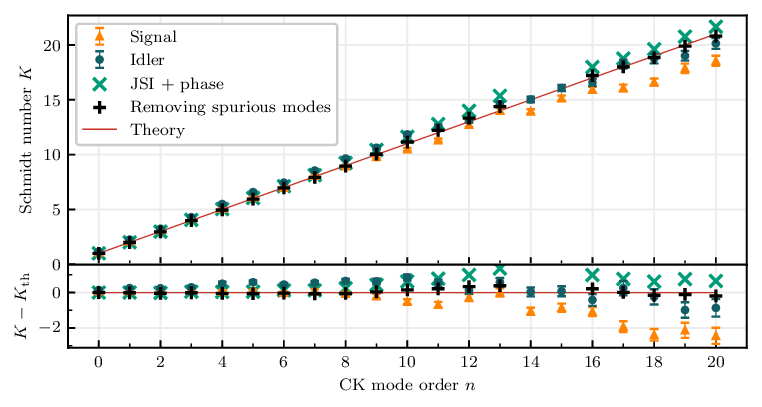}
    \caption{Experimental Schmidt number $K$ for each CK mode $n$ obtained from $g^{(2)}$ measurements in the signal and idler arms (orange and blue points), values reconstructed from the deconvolved measured JSI combined with the known pump phase (green cross), values obtained by removing the spurious modes in the Schmidt decomposition (black ``+''), compared to the theoretical values $K_\mathrm{th}$ (red line).}
    \label{fig:all-results}
\end{figure}

One can notice a very clear distinction between the first $K_\mathrm{th}$ Schmidt weights, corresponding to the modes that compose the maximally entangled state, and the following ones, which are spurious modes arising from spectral imperfections such as shaping defects or too narrow filtering, and artifacts from the grid-like structure of the pixels in the JSA. Although the weight of these spurious modes is orders of magnitude lower than that of the primary modes, their very high number introduces a small component of multi-modedness that slightly increases the value of $K$. To limit this effect, one could choose spectral filters with a bandwidth still narrow enough to cut off the side lobes of the sinc-shaped phase-matching function and to limit fluorescence, but larger than the spectral bandwidth of the generated photons. This would avoid introducing edge effects in the spectrum and would also lower the optical losses after generation. Moreover, to reach a higher level of accuracy in the generated states, one could implement an iterative optimization method to perfect the spectral shaping of the pump, which would result in a more even distribution of the Schmidt coefficients. If one were able to remove these sources of imperfections in such a way to eliminate the contributions of the spurious modes with $k>K_\mathrm{th}$, even keeping the same uneven weights of the main modes, one would obtain the effective dimensionality shown by the ``+'' symbols in Fig. \ref{fig:all-results}, which follows much more closely the theoretical predictions.

The final row of Fig. \ref{fig:jsa_rec} shows the reconstructed joint temporal intensity (JTI) calculated as the squared modulus of the Fourier transform of the reconstructed JSA. In the time picture, it becomes intuitive to see that a higher-order CK mode generates higher-dimensional entangled states in the form of correlated time bins whose intensity is determined by the Schmidt coefficients. Notably, due to the extremely short time separation between these correlated ``islands'' (approximately \SI{2}{\pico\second}), the time structure of the generated PDC states cannot be resolved by conventional single-photon detectors. For this reason, these states do not fall into the standard definition of time bins but rather correspond to that of temporal modes, which require a more sophisticated detection scheme based, e.g., on a so-called quantum pulse gate \cite{brecht14, ansari18a, serino23}.

\bibliography{bibliography_supplementary}